\documentclass[10pt,journal]{IEEEtran}
\usepackage{amsmath,amssymb,amsfonts}
\usepackage{algorithmic}
\usepackage{graphicx}
\usepackage{textcomp}
\usepackage{xcolor}
\usepackage{bm}

\begin{document}
	
\title{OFDM Enabled Over-the-Air Computation Systems with Two-Dimensional Fluid Antennas}

\author{Heyang Xiong, Quanzhong Li, and Qi Zhang, \emph{Member}, \emph{IEEE}
		
\thanks{\emph{(Corresponding author: Qi Zhang.)}
			
Heyang Xiong and Qi Zhang are with the School of Electronics and Information Technology, Sun Yat-sen University, Guangzhou 510006, China (e-mail: xionghy9@mail2.sysu.edu.cn; zhqi26@mail.sysu.edu.cn). Quanzhong Li is with the School of Computer Science and Engineering, Sun Yat-sen University, Guangzhou 510006, China (e-mail: liquanzh@mail.sysu.edu.cn).}
		
}
	
\markboth{}%
{Xiong \MakeLowercase{\textit{et al.}}: OFDM Enabled AirComp Systems}
	
\maketitle
	
\begin{abstract}
Fluid antenna system (FAS) is able to exploit spatial degrees of freedom (DoFs) in wireless channels. In this letter, to exploit spatial DoFs in frequency-selective environments, we investigate an orthogonal frequency division multiplexing enabled over-the-air computation system, where the access point is equipped with a two-dimensional FAS to enhance performance. We solve the computation mean square error (MSE) minimization problem by transforming the original problem into transmit precoders optimization problem and antenna positions optimization along with receive combiners optimization problem. The latter is solved via a majorization-minimization approach combined with sequential optimization. Numerical results confirm that the proposed scheme achieves MSE reduction over the scheme with fixed position antennas.
\end{abstract}
	
	
\section{Introduction}

With the boom of Internet-of-things and edge intelligence, a massive number of wireless devices generate huge data volumes that require efficient aggregation and processing for latency-sensitive applications. Conventional sequential, communication-based data aggregation fails to meet strict latency and resource-efficiency requirements. Over-the-air computation (AirComp) leverages the wireless signal superposition property to enable simultaneous transmission and in-air aggregation, thereby cutting communication overhead and boosting spectral efficiency \cite{GChen25,YLi22,BWei24}.

In practical wideband systems, frequency-selective fading in wireless channels introduces inter-symbol interference (ISI), degrading communication reliability. Orthogonal frequency division multiplexing (OFDM) is an extensively adopted modulation technique that mitigates ISI by dividing the wideband channel into multiple orthogonal narrowband subcarriers. The integration of AirComp with OFDM has been studied to achieve robust data aggregation in frequency-selective environments \cite{YChen24,NEvgenidis24}. 

The aforementioned works consider fixed-position antenna arrays at the access point (AP), which limits the ability to exploit spatial degrees of freedom (DoFs) for further performance enhancement. To exploit spatial DoFs, fluid antenna system (FAS) has emerged as a promising technology \cite{KKWong21,TWu24,JYao25}. By enabling continuous adjustment of antenna positions within a confined spatial region, FAS introduces additional spatial DoFs that can be optimized to improve channel conditions, such as enhancing signal-to-noise ratio (SNR), suppressing interference, and increasing spatial diversity. Recent studies demonstrate that FAS arrays can significantly reduce the mean square error (MSE) in AirComp systems by optimizing antenna positions to align with the signal superposition from multiple users \cite{DZhang24,NLi25}. 

To the best of our knowledge, the problem of exploiting spatial DoFs in frequency-selective environments remains unexplored. In this letter, we propose an OFDM enabled AirComp system enhanced by a two-dimensional (2D) FAS. Considering 
that an AP equipped with a 2D FAS aggregates the data from multiple users, we formulate the computation MSE minimization problem, which jointly optimizes AirComp-OFDM transceivers and FAS antenna positions. We decompose the original problem into two subproblems, i.e., the problem of transmit precoders optimization and that of antenna positions optimization along with receive combiners optimization. To tackle the latter optimization problem, we employ the majorization-minimization (MM) technique to derive a surrogate function and then solve it via a sequential optimization approach. 
    
\emph{Notations:} Boldface lowercase and uppercase letters denote vectors and matrices, respectively. $\mathbb{C}$ and $\mathbb{R}$ represent the sets of complex and real numbers. The transpose, Hermitian transpose, inverse, and trace of matrix $\mathbf{A}$ are denoted by $\mathbf{A}^T$, $\mathbf{A}^H$, $(\cdot)^{-1}$, and $\text{tr}(\mathbf{A})$, respectively. $\text{Re}\{a\}$ denotes the real part of a complex scalar $a$. $\|\mathbf{a}\|$ denotes the Euclidean norm of vector $\mathbf{a}$. $\text{diag}(\mathbf{a})$ denotes the diagonal matrix whose diagonal elements are given by vector $\mathbf{a}$. By $\mathbf{A}\succ\mathbf{0}$ or $\mathbf{A}\succeq\mathbf{0}$, we mean that the matrix $\mathbf{A}$ is positive definite or positive semidefinite, respectively. $\text{Blkdiag}(\mathbf{A}_1,\mathbf{A}_2,\cdots)$ constructs a matrix with the arguments $\mathbf{A}_1$, $\mathbf{A}_2$, $\cdots$ placed along its main diagonal. $\text{vec}(\mathbf{A})$ denotes the vectorization operator that stacks the columns of matrix $\mathbf{A}$ into a single column vector. $\otimes$ denotes the Kronecker product.

\section{System Model and Problem Formulation}
	
Consider an OFDM enabled AirComp system, where $K$ single-antenna users transmit data to an AP equipped with a 2D FAS consisting of $M$ antennas. 
	
The Cartesian coordinate of the $m$th FAS antenna is denoted by $\mathbf{r}_m=[x_m,y_m]^T\in\mathbb R^2$, $m\in\mathcal{M}=\{1,2,\cdots,M\}$. For notational convenience, stacking all FAS coordinates yields the antenna-position vector
\begin{equation}
\mathbf{r}=\left[\mathbf{r}_1^T,\cdots,\mathbf{r}_M^T\right]^T.
\end{equation}
The $m$th FAS antenna is constrained to lie within a movable region $\mathcal{R}$, and a minimum inter-antenna distance $\delta$ is enforced to avoid excessive mutual coupling.

Each OFDM symbol consists of $N$ subcarriers. Denote the independent information symbol transmitted by user $k$ on the $n$th subcarrier as $c_{k,n}$, which has zero mean and satisfies $\mathbb{E}[|c_{k,n}|^2]=1$. The frequency-domain precoding coefficient is denoted by $b_{k,n}\in\mathbb{C}$. The transmitted signal on the $n$th subcarrier is
\begin{equation}
d_{k,n}=b_{k,n}c_{k,n}.
\end{equation}
After applying the inverse discrete Fourier transform (IDFT), the time-domain sample transmitted by the $k$th user at time index $t$ is expressed as
\begin{equation}
\tilde{d}_{k,t} = \frac{1}{\sqrt{N}} \sum_{n=1}^{N} d_{k,n}
e^{j \frac{2\pi}{N} t n}
\end{equation}
where $t \in \{1,2,\cdots,N\}$. After that, the cyclic prefix (CP) is included before transmission through the wireless channel. 

For the $k$th user, $k\in\mathcal{K}=\{1,2,\cdots,K\}$, the channel consists of $L$ propagation paths indexed by $l\in\mathcal{L}=\{1,2,\cdots,L\}$. The $l$th path is characterized by a complex gain $g_{k,l}$, a propagation delay $p_{k,l}$, and an angle-of-arrival pair $(\theta_{k,l},\phi_{k,l})$, where $\theta_{k,l}$ and $\phi_{k,l}$ represent the elevation and azimuth angles of the $l$th receive path, respectively. Under the far-field assumption, the relative propagation distance difference of the $l$th path with respect to the $m$th antenna is approximated by
\begin{equation}
\rho_{k,l}(\mathbf{r}_m) = x_m \sin\theta_{k,l} \cos\phi_{k,l} + y_m \cos\theta_{k,l}.
\end{equation}	
The receive field-response vector corresponding to the $k$th user at the $m$th antenna is defined as
\begin{equation}\label{q7}
\mathbf{f}_k(\mathbf{r}_m) =
\left[ e^{j\frac{2\pi}{\lambda}\rho_{k,1}(\mathbf{r}_m)}, \cdots,
e^{j\frac{2\pi}{\lambda}\rho_{k,L}(\mathbf{r}_m)} \right]^T
\end{equation}
where $\lambda$ denotes the carrier wavelength. By collecting the field-response vectors associated with all antennas, the receive field-response matrix of the $k$th user is expressed as
\begin{equation}\label{q8}
\mathbf{F}_k=
\left[ \mathbf{f}_k(\mathbf{r}_1), \cdots, \mathbf{f}_k(\mathbf{r}_M) \right].
\end{equation}
The channel vector corresponding to the $l$th path between the $k$th user and the AP is given by
\begin{equation}\label{q10}
\tilde{\mathbf{h}}_{k,l}(\mathbf{r}) = g_{k,l}
\left[ e^{-j\frac{2\pi}{\lambda}\rho_{k,l}(\mathbf{r}_1)}, \cdots,
e^{-j\frac{2\pi}{\lambda}\rho_{k,l}(\mathbf{r}_M)} \right]^T.
\end{equation}
Thus, by defining
\begin{equation}\label{q11}
\tilde{\mathbf{H}}_k=\left[ \tilde{\mathbf{h}}_{k,1}(\mathbf{r}), \cdots, \tilde{\mathbf{h}}_{k,L}(\mathbf{r}) \right],
\end{equation}
we have
\begin{equation}\label{q12}
\tilde{\mathbf{H}}_k=\mathbf{F}_k^H\text{diag}(\mathbf{g}_k)
\end{equation}
where $\mathbf{g}_k=[g_{k,1},\cdots, g_{k,L}]^T$.
		
At the AP, after removing the CP, the received time-domain signal vector at time index $t$ can be written as
\begin{equation}
\tilde{\mathbf{z}}_t = \sum_{k=1}^{K} \sum_{l=1}^{L}
\tilde{\mathbf{h}}_{k,l}(\mathbf{r}) \tilde{d}_{k,t-p_{k,l}} + \tilde{\mathbf{u}}_t
\end{equation}
where $\tilde{\mathbf{u}}_t\sim\mathcal{CN}(\mathbf{0}, \sigma^2\mathbf{I})$ denotes the additive white Gaussian noise vector. By applying the discrete Fourier transform (DFT) to the received time-domain signal vector, the frequency-domain observation on the $n$th subcarrier is obtained as
\begin{equation}
\mathbf{z}_n =
\sum_{k=1}^{K} \mathbf{h}_{k,n}(\mathbf{r})
d_{k,n}+ \mathbf{u}_n
\end{equation}
where $\mathbf{u}_n\sim\mathcal{CN}(\mathbf{0}, \sigma^2\mathbf{I})$ denotes the frequency-domain noise vector and
\begin{equation}\label{q15}
\mathbf{h}_{k,n}(\mathbf{r}) =
\frac{1}{\sqrt{N}}\sum_{l=1}^{L}
\tilde{\mathbf{h}}_{k,l}(\mathbf{r})
e^{-j\frac{2\pi}{N} n p_{k,l}}.
\end{equation}

To recover the desired AirComp result, the AP applies a linear combining vector $\mathbf{w}_n \in \mathbb{C}^{M\times1}$ to estimate the target function $C_n = \sum_{k=1}^{K} c_{k,n}$ on the $n$th subcarrier. The resulting estimate is given by
\begin{equation}
\hat{C}_n = \mathbf{w}_n^H \mathbf{z}_n
= \sum_{k=1}^{K}
\mathbf{w}_n^H \mathbf{h}_{k,n}(\mathbf{r})
b_{k,n} c_{k,n}
+ \mathbf{w}_n^H \mathbf{u}_n.
\end{equation}
The corresponding MSE on the $n$th subcarrier is defined as
\begin{align}
\text{MSE}_n
&\triangleq
\mathbb{E}\left[\left|\hat{C}_n - C_n\right|^2 \right] \notag \\
&=\sum_{k=1}^{K}
\left|\mathbf{w}_n^H \mathbf{h}_{k,n}(\mathbf{r})b_{k,n} - 1\right|^2
+ \sigma^2 \|\mathbf{w}_n\|^2.
\end{align}
The overall MSE is obtained by averaging over all subcarriers,
\begin{equation}
\text{MSE} = \frac{1}{N} \sum_{n=1}^{N} \text{MSE}_n.
\end{equation}
		
Our objective is to jointly optimize the transmit precoders $\mathbf {b}_k=[b_{k,1},b_{k,2},\cdots,b_{k,N}]^T$, receive combiners $\mathbf{w}_n$, and FAS positions $\mathbf{r}$ to minimize the overall MSE, subject to power and position constraints. The resulting optimization problem is formulated as
\begin{subequations}\label{q20}
\begin{align}\label{q20a}
\min_{\mathbf{b}_{k},\mathbf{w}_n,\mathbf{r}}
&\ \mathrm{MSE}\\
\text{s.t.}\quad
&\ |b_{k,n}|^2\le P,\ \forall\ k,n,\label{q20b}\\
&\ \mathbf r_m\in\mathcal{R},\ \forall\ m,\label{q20c}\\
&\ \|\mathbf r_m-\mathbf r_{m'}\|\ge \delta,\ \forall\ m\neq m'\label{q20d}
\end{align}
\end{subequations}
where $P$ denotes the transmission power constraint on each subcarrier for every user.

\section{Iterative Optimization Algorithm}

Due to the coupled variables and the non-convex objective, problem \eqref{q20} is non-convex. We employ an alternating optimization (AO) algorithm to solve this problem.

\subsection{Optimizing $\mathbf{b}_k$ Given $\mathbf{w}_n$ and $\mathbf{r}$ }

Given $\mathbf{w}_n$ and $\mathbf{r}$, the optimization of $\mathbf{b}_k$ can be decoupled into 
$K$ independent subproblems as follows
\begin{subequations}\label{bq1}
\begin{align}\label{bq1a}
\min_{b_{k,n}}\ &\sum_{n=1}^N \left|\mathbf{w}_n^H \mathbf{h}_{k,n}(\mathbf{r})b_{k,n} - 1\right|^2 \\
\label{bq1b}\text{s.t.} \  &|b_{k,n}|^2\le P,\ \forall\ n.
\end{align}
\end{subequations}
This problem admits a closed-form solution. The optimal frequency-domain precoding coefficient on the $n$th subcarrier is given by \cite{NLi25}
\begin{equation}
b_{k,n}^o = \min\left( \sqrt{P}, \left|\mathbf{w}_n^H \mathbf{h}_{k,n}(\mathbf{r})\right|^{-1} \right) e^{-j\angle\left(\mathbf{w}_n^H \mathbf{h}_{k,n}(\mathbf{r})\right)}.
\end{equation}
		
\subsection{Optimizing $\mathbf{w}_n$ and $\mathbf {r}$ Given $\mathbf{b}_k$}

Given $b_{k,n}$ and $\mathbf{r}$, problem \eqref{q20} can be decoupled into $N$ independent subproblems as follows
\begin{equation}\label{bq6}
\min_{\mathbf{w}_n}\ \xi_n\triangleq\sum_{k=1}^K \left|\mathbf{w}_n^H \mathbf{h}_{k,n}(\mathbf{r})b_{k,n} - 1\right|^2 + \sigma^2 \|\mathbf{w}_n\|^2.
\end{equation}
Taking the first-order partial derivative of the objective function and setting it to zero gives the closed-form solution as follows
\begin{align}\label{bq8}
\mathbf{w}_n^o =&\left( \sum_{k=1}^K |b_{k,n}|^2 \mathbf{h}_{k,n}(\mathbf{r})\mathbf{h}_{k,n}(\mathbf{r})^H + \sigma^2 \mathbf{I} \right)^{-1}\nonumber\\
&\cdot\sum_{k=1}^K \mathbf{h}_{k,n}(\mathbf{r})b_{k,n}.
\end{align}
By defining
\begin{align}\label{bq10}
\mathbf{H}_n &= \left[ \mathbf{h}_{1,n}(\mathbf{r}), \cdots, \mathbf{h}_{K,n}(\mathbf{r}) \right],\\
\mathbf{B}_n &= \text{diag}\left( |b_{1,n}|^2, \cdots, |b_{K,n}|^2 \right),\\
\label{bq13}\bar{\mathbf{b}}_n &= \left[ b_{1,n},\cdots, b_{K,n} \right]^T,
\end{align}
the closed-form solution in \eqref{bq8} can be rewritten as
\begin{equation}\label{bq20}
\mathbf{w}_n^o = \left(\mathbf{H}_n \mathbf{B}_n \mathbf{H}_n^H + \sigma^2 \mathbf{I} \right)^{-1} \mathbf{H}_n \bar{\mathbf{b}}_n.
\end{equation}
Substituting \eqref{bq20} into \eqref{bq6}, the objective function of problem \eqref{bq6} is expressed as
\begin{equation}
\xi_n=K-\bar{\mathbf{b}}_n^H \mathbf{H}_n^H\mathbf{V}_n^{-1} \mathbf{H}_n \bar{\mathbf{b}}_n
\end{equation}
where 
\begin{equation}\label{bq24}
\mathbf{V}_n = \mathbf{H}_n \mathbf{B}_n \mathbf{H}_n^H + \sigma^2 \mathbf{I}.
\end{equation}
Therefore, given $\mathbf{b}_k$, problem \eqref{q20} is reduced to
\begin{align}\label{bq25}
\max_{\mathbf{r}} \ \sum_{n=1}^N \bar{\mathbf{b}}_n^H \mathbf{H}_n^H \mathbf{V}_n^{-1} \mathbf{H}_n \bar{\mathbf{b}}_n \ \ \text{s.t.} \ \eqref{q20c},\ \eqref{q20d}.
\end{align}

From \eqref{q10} and \eqref{q15}, we know
\begin{equation}\label{bq27}
\mathbf{h}_{k,n}(\mathbf{r})=\tilde{\mathbf{H}}_{k}\mathbf{e}_{k,n}
\end{equation}
where $\tilde{\mathbf{H}}_{k}$ is defined in \eqref{q11} and 
\begin{align}
\mathbf{e}_{k,n}&=\frac{1}{\sqrt{N}}
\left[ e^{-j\frac{2\pi}{N} n  p_{k,1}}, \cdots,
e^{-j\frac{2\pi}{N} n p_{k,L}} \right]^T.
\end{align}
According to \eqref{bq10}, we obtain
\begin{align}\label{bq30}
\mathbf{H}_n= \left[\tilde{\mathbf{H}}_1\mathbf{e}_{1,n},\cdots, \tilde{\mathbf{H}}_K \mathbf{e}_{K,n} \right]=
\left[\tilde{\mathbf{H}}_1,\cdots, \tilde{\mathbf{H}}_K\right]\mathbf{E}_n
\end{align}
where
\begin{equation}
\mathbf{E}_n=\text{Blkdiag}\left(\mathbf{e}_{1,n},\cdots,\mathbf{e}_{K,n}\right).
\end{equation}
Substituting \eqref{q12} into \eqref{bq30}, we obtain
\begin{align}\label{bq35}
\mathbf{H}_n=\mathbf{F}\mathbf{G}\mathbf{E}_n
\end{align}
where
\begin{align}\label{bq36}
\mathbf{F}=&\left[\mathbf{F}_{1}^H,\cdots,\mathbf{F}_{K}^H\right],\\
\mathbf{G}=&\text{diag}\left(\left[\mathbf{g}_1^T,\cdots,\mathbf{g}_K^T\right]^T\right).
\end{align}

From \eqref{bq24}, \eqref{bq27}, and \eqref{bq30}, we know problem \eqref{bq25} is non-convex because the optimization variable $\mathbf{r}$ is contained in both $\mathbf{H}_n$ and $\mathbf{V}_n$. Therefore, we propose to use the MM algorithm. In the $(\tau+1)$th iteration, assume that the optimal $\mathbf{H}_n$ and $\mathbf{V}_n$ in the $\tau$th iteration, denoted as $\mathbf{H}_n^{(\tau)}$ and $\mathbf{V}_n^{(\tau)}$, respectively, are obtained. We have the inequality
\begin{align}\label{cq1}
\text{tr}&\left(\mathbf{A}_1^H \mathbf{A}_2^{-1}\mathbf{A}_1\right)\geq2\text{Re}\left\{\text{tr}\left(\tilde{\mathbf{A}}_1^H \tilde{\mathbf{A}}_2^{-1} \mathbf{A}_1\right)\right\}\nonumber\\
&-\text{tr}\left(\tilde{\mathbf{A}}_2^{-1} \tilde{\mathbf{A}}_1 \tilde{\mathbf{A}}_1^H \tilde{\mathbf{A}}_2^{-1}\mathbf{A}_2\right)+\text{tr}\left(\tilde{\mathbf{A}}_1^H \tilde{\mathbf{A}}_2^{-1}\tilde{\mathbf{A}}_1\right)
\end{align}
where $\mathbf{A}_2\succ\mathbf{0}$ is assumed and the right-hand side is the first-order Taylor expansion of the left-hand side around $\tilde{\mathbf{A}}_1$ and $\tilde{\mathbf{A}}_2$. This inequality holds since the left-hand side of \eqref{cq1} is convex in $\mathbf{A}_1$ and $\mathbf{A}_2$ under the positive definiteness of $\mathbf{A}_2$. Using \eqref{cq1}, we obtain
\begin{align}\label{cq2}
\sum_{n=1}^N \bar{\mathbf{b}}_n^H \mathbf{H}_n^H \mathbf{V}_n^{-1} \mathbf{H}_n \bar{\mathbf{b}}_n\geq \sum_{n=1}^N\left(\Upsilon_{1,n}+ \Upsilon_{2,n} +\Upsilon_{3,n}\right)
\end{align}
where the constant term $\Upsilon_{3,n}= \bar{\mathbf{b}}_n^H (\mathbf{H}_n^{(\tau)})^H
(\mathbf{V}_n^{(\tau)})^{-1} \mathbf{H}_n^{(\tau)} \bar{\mathbf{b}}_n$ is not related with the optimization variable $\mathbf{r}$ and
\begin{align}\label{cq3}
\Upsilon_{1,n}&=2\text{Re}\left\{\bar{\mathbf{b}}_n^H \left(\mathbf{H}_n^{(\tau)}\right)^H \left(\mathbf{V}_n^{(\tau)}\right)^{-1} \mathbf{H}_n \bar{\mathbf{b}}_n\right\},\\
\label{cq4}\Upsilon_{2,n}&=- \text{tr}\left(\mathbf{S}_n^{(\tau)}\mathbf{V}_n\right),\\
\mathbf{S}_n^{(\tau)}&=\left(\mathbf{V}_n^{(\tau)}\right)^{-1} 
\mathbf{H}_n^{(\tau)} \bar{\mathbf{b}}_n \bar{\mathbf{b}}_n^H \left(\mathbf{H}_n^{(\tau)}\right)^H
\left(\mathbf{V}_n^{(\tau)}\right)^{-1}.
\end{align}
Substituting \eqref{bq13}, \eqref{bq24}, and \eqref{bq30} into \eqref{cq4}, we obtain
\begin{equation}
\Upsilon_{2,n}=-\text{tr}\left(\mathbf{F}\bm{\Lambda}_n\mathbf{F}^H\mathbf{S}_n^{(\tau)}\right)-\sigma^2\text{tr}\left(\mathbf{S}_n^{(\tau)}\right)
\end{equation}
where
\begin{align}
\bm{\Lambda}_n=\mathbf{G}\mathbf{E}_n\mathbf{B}_n\mathbf{E}_n^H\mathbf{G}^H.
\end{align}
Using the identity
\begin{equation}\label{cq10}
\text{tr}\left(\mathbf{A}_1\mathbf{A}_2\mathbf{A}_3\mathbf{A}_4\right)
=\text{vec}\left(\mathbf{A}_1^H\right)^H\left(\mathbf{A}_4^T \otimes \mathbf{A}_2\right)
\text{vec}\left( \mathbf{A}_3 \right),
\end{equation}
we have
\begin{equation}
\Upsilon_{2,n}=-\bm{\eta}^H\bm{\Psi}_n\bm{\eta}-\sigma^2\text{tr}\left(\mathbf{S}_n^{(\tau)}\right)
\end{equation}
where $\bm{\eta}=\text{vec}(\mathbf{F}^H)$ and
\begin{equation}
\bm{\Psi}_n=\left(\mathbf{S}_n^{(\tau)}\right)^T\otimes\bm{\Lambda}_n.
\end{equation}
To continue, we have the following lemma whose proof can be found in \cite{JSong15}.

\emph{Lemma 1}: Let $\mathbf{A}_1$ and $\mathbf{A}_2$ be two Hermitian matrices such that $\mathbf{A}_1\succeq \mathbf{A}_2$. Then for
any point $\mathbf{x}_0$, the quadratic function $\mathbf{x}^H\mathbf{A}_1\mathbf{x}$ is majorized
by
\begin{align}
\mathbf{x}^H\mathbf{A}_1\mathbf{x}\geq & \mathbf{x}^H\mathbf{A}_2\mathbf{x}+2\text{Re}\{\mathbf{x}^H(\mathbf{A}_1-\mathbf{A}_2)\mathbf{x}_0\}\nonumber\\ &+ \mathbf{x}_0^H(\mathbf{A}_2-\mathbf{A}_1)\mathbf{x}_0
\end{align}
at $\mathbf{x}_0$. $\hfill\blacksquare$

Using Lemma 1, by denoting the optimal $\bm{\eta}$ in the $\tau$th iteration as $\bm{\eta}^{(\tau)}$, we obtain
\begin{equation}\label{cq19}
\Upsilon_{2,n}\geq -2\text{Re}\left\{\bm{\eta}^H\left(\bm{\Psi}_n-\beta_n\mathbf{I}\right)\bm{\eta}^{(\tau)}\right\} +\kappa_n
\end{equation}
where $\beta_n$ denotes the maximum eigenvalue of $\bm{\Psi}_n$ and the constant term
\begin{equation}\label{cq20}
\kappa_n=-\beta_n\|\bm{\eta}\|^2-\left(\bm{\eta}^{(\tau)}\right)^H(\beta_n\mathbf{I}-\bm{\Psi}_n)\bm{\eta}^{(\tau)}-\sigma^2\text{tr}\left(\mathbf{S}_n^{(\tau)}\right)
\end{equation}
is not related with the optimization variable $\mathbf{r}$. In \eqref{cq20}, $\|\bm{\eta}\|^2$ is a constant because from 
\eqref{q7}, $\|\mathbf{f}_k(\mathbf{r}_m)\|^2$ is a constant.

Similarly, we rewrite $\Upsilon_{1,n}$ as
\begin{align}
\Upsilon_{1,n}&=2\text{Re}\left\{\text{tr}\left(\mathbf{H}_n\bar{\mathbf{b}}_n\bar{\mathbf{b}}_n^H \left(\mathbf{H}_n^{(\tau)}\right)^H \left(\mathbf{V}_n^{(\tau)}\right)^{-1} \right)\right\}.
\end{align}
Using \eqref{cq10}, we obtain
\begin{equation}\label{cq23}
\Upsilon_{1,n}=2\text{Re}\left\{\bm{\eta}^H\bm{\omega}_n\right\}
\end{equation}
where
\begin{equation}
\bm{\omega}_n=\text{vec}\left(\mathbf{G}\mathbf{E}_n\bar{\mathbf{b}}_n\bar{\mathbf{b}}_n^H \left(\mathbf{H}_n^{(\tau)}\right)^H \left(\mathbf{V}_n^{(\tau)}\right)^{-1}
\right).
\end{equation}

Substituting \eqref{cq19} and \eqref{cq23} into \eqref{cq2}, we have
\begin{align}\label{cq26}
\sum_{n=1}^N \bar{\mathbf{b}}_n^H \mathbf{H}_n^H \mathbf{V}_n^{-1} \mathbf{H}_n \bar{\mathbf{b}}_n\geq \sum_{n=1}^N\left(-2\text{Re}\left\{\bm{\eta}^H\bm{\psi}_n\right\}+\Upsilon_{3,n}+\kappa_n\right)
\end{align}
where
\begin{align}
\bm{\psi}_n=\left(\bm{\Psi}_n-\beta_n\mathbf{I}\right)\bm{\eta}^{(\tau)}-\bm{\omega}_n.
\end{align}

Partition $\bm{\eta}$ into $M$ disjoint blocks such that
\begin{equation}\label{cq30}
\bm{\eta}=
\left[\bm{\eta}(\mathbf{r}_1)^T, \cdots, \bm{\eta}(\mathbf{r}_M)^T\right]^T
\end{equation}
where 
\begin{equation}
\bm{\eta}(\mathbf{r}_m)=
\left[\mathbf{f}_1(\mathbf{r}_m)^T, \cdots, \mathbf{f}_K(\mathbf{r}_m)^T\right]^T
\end{equation}
for $m\in\mathcal{M}$, which is derived from \eqref{q8} and \eqref{bq36}.
	
Similarly, $\bm{\psi}_n$ can be partitioned conformably as
\begin{equation}\label{cq35}
\bm{\psi}_n=\left[\bm{\psi}_{n,1}^T, \cdots, \bm{\psi}_{n,M}^T\right]^T
\end{equation}
where $\bm{\psi}_{n,m}$ denotes the $m$th block associated with $\bm{\eta}(\mathbf{r}_m)$.
	
Substituting \eqref{cq30} and \eqref{cq35} into \eqref{cq26}, problem \eqref{bq25} is equivalently transformed into
\begin{align}\label{cq40}
\max_{\mathbf{r}} \ \sum_{m=1}^M -2\text{Re}\left\{\bm{\eta}(\mathbf{r}_m)^H\bm{\varphi}_m\right\} \ \ \text{s.t.} \ \eqref{q20c},\ \eqref{q20d}.
\end{align}
where
\begin{equation}
\bm{\varphi}_m=\sum_{n=1}^N\bm{\psi}_{n,m}.
\end{equation}
It is worth noting that the objective function of problem \eqref{cq40} is separable with respect to
$\{\mathbf{r}_m\}_{m=1}^M$, while the coupling only arises from the minimum distance constraint in \eqref{q20d}. Therefore, the $M$ positions can be optimized sequentially.		

Without loss of generality, consider optimizing $\mathbf{r}_1$ first. By ignoring $\mathbf{r}_2,\cdots,\mathbf{r}_M$, problem \eqref{cq40} is reduced to
\begin{equation}
\max_{\mathbf{r}_1}\ \ -2\text{Re}\left\{\bm{\eta}(\mathbf{r}_1)^H\bm{\varphi}_1\right\}
\ \text{s.t.}\ \mathbf{r}_1 \in \mathcal{R},
\end{equation}
which can be solved by a two-dimensional search over $\mathcal{R}$.
		
Denote the optimal solution as $\mathbf{r}_1^o$. Given $\mathbf{r}_1^o$, the feasible region for $\mathbf{r}_2$ is reduced by excluding a circular area centered at $\mathbf{r}_1^o$ with radius $\delta$, denoted as $\mathcal{D}(\mathbf{r}_1^o,\delta)$, as required by the minimum distance constraint \eqref{q20d}. Accordingly, the optimization of $\mathbf{r}_2$ is given by
\begin{equation}
\max_{\mathbf{r}_2}\ \ -2\text{Re}\left\{\bm{\eta}(\mathbf{r}_2)^H\bm{\varphi}_2\right\}
\ \text{s.t.}\ \mathbf{r}_2 \in \mathcal{R}\setminus \mathcal{D}(\mathbf{r}_1^o,\delta).
\end{equation}
This procedure is repeated sequentially. At the $m$th step, the feasible region for $\mathbf{r}_m$ is 
\begin{equation}
\mathcal{R}\setminus \left(\mathcal{D}(\mathbf{r}_1^o,\delta)\cup\cdots\cup\mathcal{D}(\mathbf{r}_{m-1}^o,\delta)\right).
\end{equation}

\section{Numerical Results}

In simulations, we assume that in the OFDM enabled AirComp system with 2D FAS, the carrier frequency is 2.4 GHz, corresponding to a wavelength of $\lambda=0.125$ m. If not specified, the number of users is $K=5$. The number of subcarriers is $N=64$. The FAS consists of $M=4$ antennas \cite{JYao252}. The minimum inter-antenna distance is $\delta=\lambda/2$. The feasible region for FAS is $\mathcal{R}$ defined as that over $-3\lambda/2\leq x_m\leq 3\lambda/2$ and  $-3\lambda/2\leq y_m\leq 3\lambda/2$ for $m\in\mathcal{M}$. The channel consists of $L=4$ propagation paths. The angle-of-arrival pair $\theta_{k,l}$ and $\phi_{k,l}$ are uniformly distributed over $[0, \pi)$ for $k\in\mathcal{K}$ and $l\in\mathcal{L}$. The complex path gain $g_{k,l}\sim \mathcal{CN}(0,1/L)$ for $k\in\mathcal{K}$ and $l\in\mathcal{L}$. 

In Fig. 1, we compare the computation MSE of our proposed OFDM enabled AirComp system with 2D FAS, denoted as ``Proposed" with other schemes for different values of $P/\sigma^2$. In the legend, ``FPA" denotes the scheme with fixed position antennas where all antennas equipped at the AP are deployed at fixed positions whose Cartesian coordinates are $\mathbf{r}_1=[-3\lambda/2,0]^T$, $\mathbf{r}_2=[-\lambda/2,0]^T$, $\mathbf{r}_3=[\lambda/2,0]^T$, and $\mathbf{r}_4=[3\lambda/2,0]^T$. ``EAS" denotes the exhaustive search scheme where $M$ antennas are selected from $2M$ positions with $x_m=m_1\lambda/2$, $m_1\in\{-3, -1, 1, 3\}$ and $y_m=m_2\lambda/2$, $m_2\in\{-1, 1\}$. ``SCA" refers to the use of the successive convex approximation algorithm proposed in \cite{JYao252} for solving problem \eqref{cq40}. From Fig. 1, it is observed that our proposed scheme outperforms the ``FPA", ``EAS", and ``SCA" schemes in terms of
MSE reduction across the entire $P/\sigma^2$ range. Furthermore, as $P/\sigma^2$ increases, the performance gap between our proposed scheme and other schemes widens.

\begin{figure}
\centering
\includegraphics[width=3.6in]{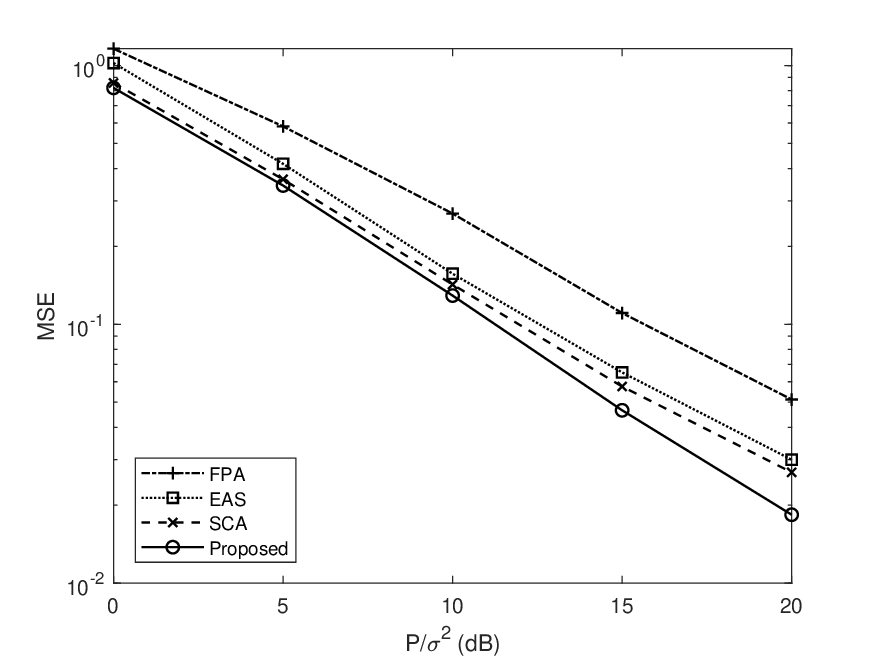}
\caption{MSE versus $P/\sigma^2$; comparison of different schemes, where $K = 5$, $N = 64$, and $M=4$.}
\end{figure}

In Fig. 2, we compare the computation MSE of different schemes for different values of $K$, where $P/\sigma^2=10$		dB. It is found from Fig. 2 that the MSE values of all schemes rise with the increase of $K$. This shows that a larger $K$ leads to lower AirComp accuracy. Furthermore, our proposed scheme yields the lowest MSE values across all tested values of $K$.

\begin{figure}
\centering
\includegraphics[width=3.6in]{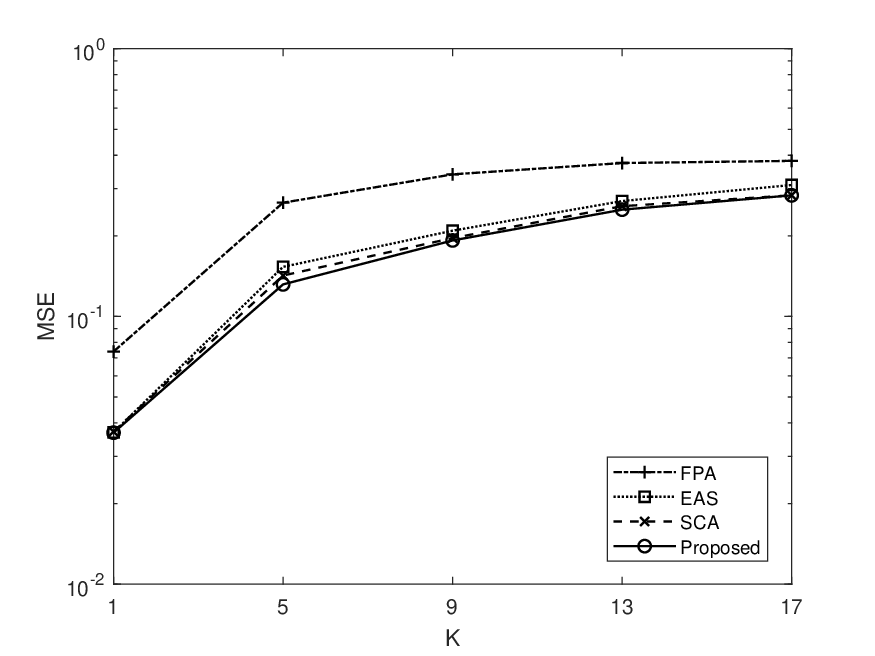}
\caption{MSE versus $K$; comparison of different schemes, where $P/\sigma^2=10$ dB, $N = 64$, and $M=4$.}
\end{figure}

\section{Conclusion}

In this letter, we have proposed a joint transceiver and antenna position optimization scheme for an OFDM enabled AirComp
system enhanced by 2D FAS. To minimize the computation MSE, we have also derived an AO algorithm combined with the MM approach, in which the closed-form transmit precoders and the antenna positions optimized via sequential two-dimensional search are updated alternately. It is shown through numerical results that the proposed OFDM enabled AirComp with 2D FAS scheme is superior to the scheme with fixed position antennas.


\begin{thebibliography}{1}
\bibitem{GChen25} G. Chen, J. Li, Q. Wu, M. Hua, K. Meng, and Z. Lyu, ``Intelligent reflecting surface aided AirComp: Multi-timescale design and performance analysis," \emph{IEEE Trans. Veh. Technol.}, vol. 74, no. 4, pp. 6644-6649, Apr. 2025.

\bibitem{YLi22} Y. Li, M. Jiang, G. Zhang, and M. Cui, ``Joint optimization for multiantenna AF-relay aided over-the-air computation," \emph{IEEE Trans. Veh. Technol.}, vol. 71, no. 6, pp. 6744–6749, Jun. 2022.
    
\bibitem{BWei24} B. Wei, P. Zhang, and Q. Zhang, ``Active reconfigurable intelligent surface-aided over-the-air computation networks," \emph{IEEE Wireless Commun. Lett.}, vol. 13, no. 4, pp. 1148-1152, Apr. 2024.

\bibitem{YChen24} Y. Chen, H. Xing, J. Xu, L. Xu, and S. Cui, ``Over-the-Air Computation in OFDM Systems with imperfect channel state information," \emph{IEEE Trans. Commun.}, vol. 72, no. 5, pp. 2929-2944, May 2024.
    
\bibitem{NEvgenidis24} N. G. Evgenidis, S. A. Tegos, P. D. Diamantoulakis, and G. K. Karagiannidis, ``Over-the-air computing in OFDM Systems," \emph{IEEE Commun. Lett.}, vol. 28, no. 11, pp. 2523-2527, Nov. 2024.
    
\bibitem{KKWong21} K.-K. Wong, A. Shojaeifard, K.-F. Tong, and Y. Zhang, ``Fluid antenna systems," \emph{IEEE Trans. Wireless Commun.}, vol. 20, no. 3, pp. 1950-1962, Mar. 2021.
    
\bibitem{TWu24} T. Wu et al., ``Fluid antenna systems enabling 6G: Principles, applications, and research directions," 2024, \emph{arXiv:2412.03839.} [Online]. Available: http://arxiv.org/abs/2412.03839

\bibitem{JYao25} J. Yao, T. Wu, L. Zhou, M. Jin, C. Huang, and C. Yuen, ``FAS versus ARIS: Which is more important for FAS-ARIS communication systems?" \emph{IEEE Trans. Wireless Commun.}, vol. 25, pp. 2075-2091, 2026.

\bibitem{DZhang24} D. Zhang, S. Ye, M. Xiao, K. Wang, M. Di Renzo, and M. Skoglund, ``Fluid antenna array enhanced over-the-air computation," \emph{IEEE Wireless Commun. Lett.}, vol. 13, no. 6, pp. 1541-1545, Jun. 2024.

\bibitem{NLi25} N. Li, P. Wu, B. Ning, L. Zhu, and W. Mei, ``Over-the-air computation via 2-D movable antenna array," \emph{IEEE Wireless Commun. Lett.}, vol. 14, no. 1, pp. 33-37, Jan. 2025.

\bibitem{JSong15} J. Song, P. Babu, and D. P. Palomar, ``Optimization methods for designing sequences with low autocorrelation sidelobes," \emph{IEEE Trans. Signal Process.}, vol. 63, no. 15, pp. 3998-4009, Aug. 2015. 

\bibitem {JYao252} J. Yao et al., ``FAS-driven spectrum sensing for cognitive radio networks," \emph{IEEE Internet Things J.}, vol. 12, no. 5, pp. 6046-6049, Mar. 2025.
\end{thebibliography}
\end{document}